\documentclass[10pt,twocolumn,letterpaper]{article}
\usepackage{cvpr}
\usepackage{times}
\usepackage{epsfig}
\usepackage{graphicx}
\usepackage{amsmath}
\usepackage{amssymb}
\usepackage{algorithm}
\usepackage{algpseudocode}
\usepackage{multicol}
\usepackage{booktabs}
\usepackage[table]{xcolor}
\usepackage{mathtools}

\usepackage[pagebackref=true,breaklinks=true,letterpaper=true,colorlinks,bookmarks=false]{hyperref}

\cvprfinalcopy 

\def\cvprPaperID{****} 

\ifcvprfinal\fi

\newcommand{\Ltwo}{L^2([0,2\pi],\mathbb{R}^3)}
\newcommand{\ltwo}{L^2}
\newcommand{\cl}{\mathbb{R}}
\newcommand{\Tms}{T_{\mu}S}
\newcommand{\Om}{\Omega}
\newcommand{\Or}{\mathcal{O}}
\newcommand{\Pb}{\mathcal{P}}
\newcommand{\Sp}{\mathcal{C}^{o}}

\newcommand{\Diff}{\mathcal{D}}
\newcommand{\inner}[2]{\left\langle #1,#2 \right\rangle}
\def\argmin{\mathop{\rm argmin}}

\begin{document}

\title{A Riemannian Framework for Linear and Quadratic Discriminant \\ Analysis on the Tangent Space of Shapes}

\author{
Susovan Pal,  Roger P. Woods,\\
UCLA Brain Mapping Center\\
University of California, Los Angeles\\
Los Angeles, CA, USA\\
{\tt\small susovanpal@mednet.ucla.edu, rwoods@ucla.edu}
\and
Suchit Panjiyar\\
Department of Computer Science\\
University of California, Los Angeles\\
Los Angeles, CA, USA\\
{\tt\small suchit@ucla.edu}
\and
Elizabeth Sowell\\
Department of Pediatrics\\
Children's Hospital Los Angeles\\
University of Southern California, Los Angeles\\
Los Angeles, CA, USA\\
{\tt\small esowell@chla.usc.edu}
\and
Katherine L. Narr, Shantanu H. Joshi \\
UCLA Brain Mapping Center\\
University of California, Los Angeles\\
Los Angeles, CA, USA\\
{\tt\small narr@uclae.edu, s.joshi@ucla.edu}
}

\maketitle

\begin{abstract}

We present a Riemannian framework for linear and quadratic discriminant classification on the tangent plane of the shape space of curves. The shape space is infinite dimensional and is constructed out of square root velocity functions of curves. We introduce the idea of mean and covariance of shape-valued random variables and samples from a tangent space to the pre-shape space (invariant to translation and scaling) and then extend it to the full shape space (rotational invariance). The shape observations from the population are approximated by coefficients of a Fourier basis of the tangent space. The algorithms for linear and quadratic discriminant analysis are then defined using reduced dimensional features obtained by projecting the original shape observations on to the truncated Fourier basis. We show classification results on synthetic data and shapes of cortical sulci, corpus callosum curves, as well as facial midline curve profiles from patients with fetal alcohol syndrome (FAS).

\end{abstract}


\section{Introduction}

Often, anatomical features can be compactly represented by  curve-based landmarks or boundaries. Depending on the study in question, this practice can have several advantages;  i) it is easier for anatomical experts to delineate or identify such curve based features in images or even on $3D$ parametrized representations of images (surfaces etc.), ii) it is convenient for computational analysis since there is a large reduction of the feature space that contributes to increased numerical efficiency of representation and analysis, and iii) if the features are easy to identify and trace, they can be quickly used for a first level analysis to determine changes in anatomy due to disease or development. 

Briefly, such biological curve-shape based studies  are outlined as follows. The white matter morphology of the mid-sagittal cross-section of the corpus callosum in the brain can represented by a  boundary curve whose shape has been implicated in various diseases. The callosal morphology shows abnormalities in  fetal alcohol syndrome \cite{Bookstein:2002eo,Sowell:2001va} and in relation to  autism spectrum disorders \cite{wolff2015altered}. It has also been implicated in schizophrenia  \cite{joshi2013statistical} and personality disorders \cite{downhill2000shape}. On the other hand,  sulcal curves on the brain are examples of cortical landmarks \cite{joshi2010cortical}. With a well established protocol, major sulcal and gyral features (12 $\sim$13 sulci) can be consistently delineated by neuroanatomists \cite{joshi2012diffeomorphic}.  There are other examples of curve shape analysis from biological imaging -- cellular shapes  from microscopy \cite{an2012modeling,ambuhl2012high} and echocardial curves from  ultrasound images \cite{joshi2009intrinsic}.  

Most of the above works have explicitly used curve representations of biological objects without  focusing on the geometry of the underlying space of such curves, although Wolff et al. \cite{wolff2015altered} use an invariant medial shape representation that can be further used to construct geometric shape spaces. However for most approaches there is a mismatch of the cost function for shape matching and the underlying metric of the space to which these curves belong. Further, most studies have largely focused on population level group differences based on point-wise statistical tests conducted directly on the matched curves. A recent approach by Bogunovic, Frangi et al. \cite{bogunovic2012automated} have used the well known LDDMM framework for the characterization of internal carotid artery curves. They obtained a $70$ to $80\%$ cross validation classification rate based on global features (curvatures and torsion ratios etc.) of the curves.  While we recognize that curve features should be best tuned according to the application at hand, in this work we adopt a different approach. Our goal here is to construct a classification framework using the full shape of the curves by following a differential geometric approach that exploits the geometry of the shape space. Specifically, in this paper we construct a Fourier basis on the tangent space of the shape space and approximate the shape observations by their tangent vectors parametrized by the basis. In this paper we focus on two simple classifiers -- the linear and the quadratic discriminant classifiers, although the framework allows the definition of more advanced classifiers as well. 

This paper is organized as follows. Section \ref{sec:srvf} briefly outlines our choice of the curve representation. We use the square root velocity functions (SRVF)  \cite{Joshi:2007p1253,joshi2007removing,srivastava2011shape} to define shapes of curves.  We chose the square root velocity formulation for curves as it is an explicit representation and simple to compute. 
Sections \ref{sec:mean_preshape} and \ref{sec:mean_shape} outline our notion of the mean and covariance on the pre-shape and the shape space respectively. An important ingredient for approximating the tangent space of shapes is the construction of a Fourier basis \ref{sec:fourier},  which is used to approximate the tangent space. Finally, section \ref{sec:classifier} defines the classifier on the tangent space followed by results in section \ref{sec:results} on synthetic data as well as on data with clinical relevance.   The novelty of our work has two fronts; a tangent space classification approach  defined for this particular geometric representation of curves, and  the first attempt to classify fetal alcohol syndrome using the midline curves of $3D$ facial images. To our knowledge this has not been previously tried or accomplished. 

\section{Brief background of shape representation}

We briefly introduce the square-root velocity function (SRVF), preshape and shape spaces of curves here. For a detailed introduction, please see \cite{Joshi:2007p1253,srivastava2011shape}. 

\subsection{Square Root Velocity representation and Preshape spaces of Parametrized curves}
\label{sec:srvf}

Let $\beta$ be a parametrized curve $\beta: D:=[0,2\pi] \to \cl^{n}$. We will restrict to
those $\beta$ that are almost everywhere differentiable and their first derivative is in
$\ltwo(D,\cl^n)$. For the purpose of
studying the shape of $\beta$, we will represent it using the square-root
velocity function (SRVF) \cite{Joshi:2007p1253,srivastava2011shape} defined as $q: D \to \cl^n$, where $ q(t):= \dot{\beta}(t) / \sqrt{ \|  \dot{\beta(t)} \| }$. This
representation includes those curves whose parametrization can become
singular in the analysis. Also, for every $q \in \ltwo(D,\cl^n)$ there
exists a curve $\beta$ (unique up to a translation) such that the given $q$
is the SRV function of that $\beta$, given by, $\beta(t)  = \int_0^t q(s) \|q(s)\| ds$.  To remove scaling variability, we re-scale all curves to be of length
$2\pi$. The remaining transformations (rotation, translation, and
re-parametrization)  will be dealt with differently. The restriction that
$\beta$ is of length $2\pi$ translates to the condition that $\int_{D}
\|q(t)\|^2 dt = \int_{D} \|\dot{\beta}(t) \| dt = 2\pi$. Therefore, the SRV
functions associated with these curves are elements of the unit sphere in the
Hilbert space $\ltwo(D,\cl^{n})$; we will use the notation ${\cal C}^o$
to denote this hypersphere.

We will call ${\cal {C} }^o$ the pre-shape space of curves. To
impose Riemannian structures on this pre-shape space, we consider their
tangent spaces. Since ${\cal C}^o$ is a sphere in $\ltwo([0,2\pi],\cl^n)$,
its tangent space at a point $q$ is given by:
$T_q({\cal C}^o) = \{v\in \ltwo([0,2\pi],\cl^{n}) | \langle v,q \rangle = 0\}.$
Here $ \langle v,q \rangle$ denotes the inner product in
$\ltwo([0,2\pi],\cl^n)$: $ \langle v,q \rangle = \int_0^{2\pi}
(v(t).q(t))_{\cl^n} dt$.\\

The standard metric on $\ltwo([0,2\pi],\cl^n)$ restricts to 
one on ${\cal C}^o$. This metric can then be used to determine geodesics and geodesic lengths
between elements of these spaces, which is straightforward, ${\cal C}^o$ being a hypersphere owing to the global scaling constraint.

\subsection{Shape Spaces of Parametrized Curves}
\label{sec:preshape}
By representing a parametrized curve $\beta(t)$ by its SRV function
$q(t)$, and imposing the constraint $\int_D \inner{q(t)}{q(t)} dt = 2\pi$,
we have taken care of the translation and the scaling variabilities, but the
rotation and the re-parametrization variabilities still remain. A rotation
is an element of $SO(n)$, the special orthogonal group of $n \times n$
matrices, and a re-parametrization is an element of $\Gamma$, the set of
all orientation-preserving diffeomorphisms of $D$.\\
The rotation and re-parameterization of a curve $\beta$ are denoted by the
actions of $SO(n)$ and $\Gamma$ on its SRV. While the action of $SO(n)$ is
the usual: $SO(n) \times {{\cal C}^o} \to { {\cal C}^o },\ \ \ (O,q(t)) = Oq(t)$, the
action of $\Gamma$  is derived as follows. For a $\gamma \in \Gamma$, the
composition $\beta \circ \gamma$ denotes its re-parameterization; the SRV
of the re-parametrized curve is $ F( \dot{\beta}(\gamma(t))
\dot{\gamma}(t)) = q(\gamma(t)) \sqrt{\dot{\gamma(t)}}$, where $q$ is the
SRV of $\beta$. This gives use the action $\Gamma \times {{{\cal C}^o}} \to {{\cal C}^o},\ \ \ (q, \gamma) = (q \circ \gamma) \sqrt{\dot{\gamma}}$. It can be shown that i)  the actions of  $SO(n)$ and $\Gamma$ on ${\cal C}^o$ commute, and that ii) the action of the product group $\Gamma \times
SO(n)$ on ${{{{\cal C}^o}}}$ is by isometries with respect to the chosen metric.\\
Therefore, under the actions of $O$ and $\Gamma$, we can define the quotient space of ${\cal C}^o$ modulo $\Gamma
\times SO(n)$. The orbit of a function $q \in {\cal C}^o$ is given by:
$[q] =  \{O(q \circ \gamma)\sqrt{\dot{\gamma}}) | (\gamma,O) \in \Gamma \times SO(n)\}.$
In this framework, an orbit is associated with a shape and comparisons
between shapes are performed by comparing the orbits of the corresponding
curves and, thus, the need for a metric on the set of orbits. In order for
this set to inherit the metric from ${\cal C}^o$, we need the orbits to be
closed sets in ${\cal C}^o$. Since these orbits are not closed in ${\cal C}^o$,
we replace them by their closures in $\ltwo(D,\cl^n)$. With a slight
abuse of notation, we will call these orbits $[q]$. Then,  the
quotient space ${\cal S}$ is defined as the set of all such closed orbits associated
with the elements of ${\cal C}^o$, i.e. ${\cal S} = \{ [ q ] | q \in {\cal C}^o \}$. The differential of this quotient map from ${\cal C}^o$ to ${\cal S}$ induces a linear isomorphism between $T_{[q]}({\cal S})$ and, the normal space to $[q]$ at any point $\tilde{q}$ in $[q]$. The Riemannian
metric on ${\cal C}^o$ restricts to an inner
product on the normal space which, in turn, induces an inner product on
$T_{[q]}({\cal S})$. The fact that $\Gamma \times SO(n)$ act by isometries
implies that the resulting inner product on $T_{[q]}({\cal S})$ is
independent of the choice of $\tilde{q} \in [q]$. In this manner, ${\cal
S}$ inherits a Riemannian structure from ${\cal C}^o$. See \cite{Joshi:2007p1253,srivastava2011shape} for computation of geodesics in ${\cal
S}$.

\section{Statistics on vector subspaces of $\Ltwo$}
\label{sec:mean_preshape}

From this section onwards, for practical purposes, we will assume that $n=3.$
Let $V \subset \Ltwo$ be a subspace of the Hilbert space $\Ltwo$. We will see in the next section that $V$ will be mainly $\Tms$ for this paper, which has been shown to be a subspace of $\Ltwo$ \cite{Joshi:2007p1253,srivastava2011shape}. Here, we introduce the concepts of mean and covariance for $V$-valued data, and their discretized versions for computer implementation. This is done keeping in mind that, in the future, the pre-shape data will be represented by elements in $\Ltwo$, and their tangent space will be represented by $V$, so we can carry the concepts in this section over to shape data. Although the shape curve data is typically continuous, being elements of $\Ltwo$, for the purpose of  computer implementation, we deal with discretized versions of the shape data, where each datum is considered to be an ordered set of points on elements of $\Ltwo$. With this in mind, we define three related concepts of mean and covariance: i) mean and covariance of a random variable $ X: \Omega \to V \subset \Ltwo $, ii) mean and covariance of a sample of  parameterized curves $\{v_1, ...v_N\} \subset V $ generated by the random variable $X$, and finally, iii) mean and covariance of discretized versions of the curves $\{v_1, ...v_N\}$, where each $v_j$ is measured at $m$ equispaced time intervals dividing $[0,2\pi]$. This is done for the purpose of discretization as  we only consider finite number of equispaced points on the curves $\{v_1, ...v_N\} \subset \Ltwo.$

\subsection{Mean and covariance of a $V \subset \Ltwo$-valued random variable:}
Let $X: \Omega \to V \subset \Ltwo$ be a random variable with probability measure $\Pb$ on $\Om$. Choose an orthonormal Schauder basis (we will interchangeably call it a Fourier basis) $W$ for $\{w_1, w_2, ...w_k, ...\}$ for $V$. Let $X(\omega)=\sum_{j=1}^{\infty}X_j(\omega)w_j \forall \omega \in \Omega$, where each $X_j=\langle X,w_j \rangle_{V}$ is a real-valued random variable, where $ \langle , \rangle_V$ is the induced inner product on $V$ from $\Ltwo$. We define mean or expected value of X, denoted by $E(X)$ as $E(X):=\sum_{j=1}^{\infty}E(X_j)w_j \in V$. Next, we define covariance of $X$ by the linear operator on $V$ given by: $cov(X)(w_j):=\sum_{k=1}^{\infty}cov(X_j, X_k)w_k$, and then linearly extending it. Hence, with respect to the orthonormal Schauder basis $W$, we have the infinite matrix representation for $cov(X)$ given by $cov(X)(i,j):= cov(X_i, X_j) = \mbox{cov} ( \langle X, w_i \rangle, \langle X, w_j \rangle)$.  

\subsubsection{Motivation for using a Fourier basis}
\label{sec:motivation_fourier}

A Fourier basis for $V$ is, by definiton, a Schauder basis \cite{kufner1977function}, which is also orthonormal. This orthonormality makes the representation of covariance much easier. The covariance matrix for a random vector $X=(X_1, X_2,...X_d) : \Omega \to \cl^d$ has the $(i,j)$-entry $\mbox{cov} ( X_i, X_j) \coloneqq \mbox{cov} ( \langle X, e_i \rangle, \langle X, e_j \rangle)$. Here $e_i$ denotes the orthonormal basis element $(0,..1...)$ with $1$ in the $i$-th slot. This matrix can also be identified by the  linear map on $\cl^d$ and denoted by $\mbox{cov}(X)$, so that $\mbox{cov}(X) (e_i) \coloneqq  \sum_{j} \mbox{cov}( \langle X, e_i \rangle, \langle X, e_j \rangle) e_j$. Note that the above definition of covariance uses the pairwise covariance between $X_i:=\langle X, e_i \rangle$. There are two motivating criteria for choosing a Fourier basis. For an arbitrary vector space $V$ with an orthonormal Fourier basis $\{w_i\}$, the above definition of covariance as an operator on $V$ is realized by replacing $e_i$ by $w_i$. This is explained in detail in the next subsection (Sec. \ref{sec:mean_discrete}). Secondly and importantly, the lower order harmonics of the Fourier basis capture the gradual changes, whereas higher order terms capture details and minutiae. Restricting the basis to finite lower order terms also enables a natural smoothing of functions in the underlying space, which makes the Fourier basis an attractive choice.

\subsection{Mean and covariance of a finite discrete sample lying in $V \subset \Ltwo$:}
\label{sec:mean_discrete}
Let $\{v_1, ...v_n\}$ be a sample generated in that order by the above random variable $X: \Omega \to V \subset \Ltwo$. We define the sample mean by $m:=\frac{1}{n}\sum_{i=1}^{n}v_i$, where the addition takes place in the vector space $V$. One can see the connection between the sample mean and the (population) mean or expectation of a random variable $X$ as follows: divide the sample space $\Omega$ into $n$ number of sub-sample spaces $\{ \Omega_1, ...\Omega_n \}$ so that $\Pb(\Omega_j)=\frac{1}{n} \ \forall 1\leq j \leq n $, assume that $X$ takes constant values $v_j$ on $\Omega_j$, and then calculate its expected value $E(X)=\int_{\Omega}Xd{\Pb}=\sum_{j=1}^{n}\int_{\Omega_j}v_jd{\Pb}=\sum_{j=1}^{n}v_j\int_{\Omega_j}d{\Pb}=\sum_{j=1}^{n}v_j. \Pb(\Omega_j)=\frac{1}{n}\sum_{j=1}^{n}v_j$, since $\Pb(\Omega_j)=\frac{1}{n} \forall 1\leq j \leq n $.\\

\noindent Before defining the sample covariance in $V$, we revisit the definition of the same in standard $\cl^d$-valued dimensional data with $d$ features. Recall that, in order to define covariance of a sample of $n$ observations $\{v_1, v_2 ...,v_n\}$, each with $d$ number of features, we write the (ordered) data matrix in $X$ two ways: $X=[v_1, v_2, ...v_j...v_n]=[R_1,...R_i...R_d]^{T}$, taking the $j$-th observation $v_j$ as its $j$-th column, and $i$-th feature $R_i$ as its $i$-th row.. Let $\bar{X}:= (\bar{R_1},...\bar{R_d})^{T} \in \cl^{d \times 1}$, where $\bar{R_i}:=\frac{1}{n}R_{ij}$. Note that, $\bar{X}$ is nothing but the sample mean $m$. Let $\tilde{X}:=\bar{X}(1, 1, ...1)\in \cl^{d \times n}$. Next, we define the (unbiased) covariance matrix to be the $d \times d$ matrix $cov(X):=\frac{1}{n-1}(X-\tilde{X})(X-\tilde{X})^{T}$.

\noindent Now, to generalize the above definition to the case of $V$-valued data, we need to be careful: firstly, $V$ may not be a finite dimensional vector space, and secondly, $V$ does not have the canonical inner product structure of $\cl^d$ giving orthogonal canonical directions of the form $(0,...0,1,0,...0)$, hence the feature directions may be unclear and may depend on the basis chosen.  The first difficulty cannot be theoretically overcome as the covariance matrix must have the dimensions corresponding to the number of linearly non-redundant features. We attempt  to solve the second difficulty simultaneously by defining features so that the feature directions (but not the feature values) are mutually orthogonal, and so that corresponding feature values are the projection of each shape data into the corresponding feature's direction. An obvious choice is to express each $a\in V$ by $a=\sum_{j=1}^{\infty}a_j w_j$, declare $w_j \in V$ as the $j$-th feature direction of $a$, and the corresponding feature value to be the $j$-th projection $a_j:= \langle v,w_j \rangle_{V}$. This way, the data matrix $X$ of the ordered sample $[v_1, v_2, ...v_j...v_n]$, as written in the previous paragraph, becomes an $\infty$ by $n$ matrix with respect to the orthonormal Schauder or Fourier basis $W=\{w_1, w_2, ...w_i, ...\}$ of $V$, and is written in two ways: $X=[C_1, C_2, ...C_j...C_n]= [R_1, R_2, ...R_i...]^{T}$, where the $j$-th column $C_j$ is so that its $i$-th entry, or equivalently the $(i,j)$-th data matrix entry $R_{ij}$ is $ \langle v_j, w_i \rangle_{V}$. Next, we define $\bar{R_i}:=\frac{1}{n}\Sigma_{j=1}^{n} R_{ij} \forall 1\leq i \leq \infty$, $\bar{X}:= [\bar{R_1},...\bar{R_{\infty}}]^{T} \in \cl^{\infty \times 1}$, and $\tilde{X}:=\bar{X}(1, 1, ...1)\in \cl^{\infty \times n}$. Then, we can \textit{represent} $cov(X): V\to V$ in this Fourier basis by the $\infty$ by $\infty$ matrix given by $cov(X)=\frac{1}{n-1}(X-\tilde{X})(X-\tilde{X})^{T}.$

\subsection{Mean and covariance of a finite set of sample curves in $V\subset\Ltwo$ discretized at $(m+1)$ points $t_r,0\leq r \leq m:$}
\label{sec:mean_finite}
So far, the elements of $V \subset \Ltwo$ have been treated as continuous curves. For example, in the previous paragraph,  $\langle v, w_j \rangle_{V}$ is the standard Euclidean inner product between the vector valued functions $v$ and $w$.  For real data we would only have discretized values of the sample $v$, discretized say at $(m+1)$ time points $t_r, 0 \leq r \leq m$ on the interval $[0,1]$. We are also going to assume here that $t_r=r/m, 0 \leq r \leq m$. With this assumption, the vector $v_j \in V$ becomes the infinite column vector $C_j$ as above, but the $k$-th entry of $C_j$ or equivalently the data matrix entry $X_{jk}$ is no more $\langle v_j, w_k \rangle_{V}=\int_{0}^{1} \langle v_j(t), w_k(t) \rangle_{{\cl}^3} dt$, but rather its discrete version $\sum_{r=0}^{m} \langle v_j(t_r), w_k(t_r) \rangle_{{\cl}^3}$. Choose a $K$ big enough, and hence $X$ becomes a $K \times n$ matrix, where one can think of this $K$ as the first $K$ most ``prominent'' features. This completes the computation of the data matrix $X$, and subsequently the $K \times K$ covariance matrix $\frac{1}{n-1}(X-\tilde{X})(X-\tilde{X})^{T}$.

\section{Measures of central tendency and deviation on Curve Shapes}
\label{sec:mean_shape}

In this section, we describe how to compute the means and covariances of data lying on shape manifold that occurs in square root velocity framework \cite{Joshi:2007p1253,srivastava2011shape}.  But first we briefly remind the reader of the calculation of the population mean of a sample of shapes in ${\cal S}$. Given a set of observations $\{ q_i \}, i = 1, \ldots, N$, their K{\"a}rcher mean \cite{karcher1977riemannian}  is  defined as 
\begin{equation}
\mu=\argmin_{q} \frac{1} {N} \sum_{i=1}^{N}  \argmin_{O, \gamma} d (q , \sqrt{\dot{\gamma}} \ O (q_{i} \circ \gamma))^{2},i=1,\ldots,N,
\label{eq:karcher}
\end{equation}
where $d(, )$ is given by the geodesic distance (great circle lengths) on ${\cal C}^o$.  This geodesic is given by the exponential map $\mbox{exp}_{q_1}(f) = q_1 \cos \alpha + f \sin \alpha$, where $cos \ \alpha = \langle q_1, q_2\rangle$ and the initial tangent vector is given by $f=q_2 - \langle q_1, q_2\rangle$. 


One standard way is to construct and better understand the elements of the tangent spaces $V=\Tms$ of the shape spaces, and replicate in the vector space $V$ the mean and covariance calculations as described in the previous two subsections. This calculation will be valid, of course if we can show that $V$ is indeed a subspace of $\Ltwo$, and find an orthonormal Schauder or Fourier basis for it.

\subsection{Construction of a Fourier basis on the tangent space of the shape space}
\label{sec:fourier}
To describe the elements of $\Tms$, we first note that we have the following orthogonal direct sum: $T_{q}{\Sp}=\Tms \oplus T_q{\Or(q)}$, where the preshape $q \in \Sp$ is a point of its orbit $\Or(q)$ so that $\mu$ is the equivalence class $[q]$ of $q$, under the action of $SO(3) \times \mathcal{D}$. Also, since $\Sp$ is a hypersphere in $\Ltwo$, it is  also clear that $T_{q}\Sp = {q}^{\perp}\subset \Ltwo$. Hence in order to construct a basis for $\Tms$, we first focus on doing the same for $T_q{\mathcal{O}(q)}$, the computation of which follows shortly after. Once this is done, it is straightforward to construct a basis for $\Tms$, since the construction of a basis for $T_{q}{\Sp={q}^{\perp}}$ is easy,  from  that of $\Ltwo$.\\

\noindent Below we present a general form of an element in $T_q{\mathcal{O}(q)}$, which in turn helps us to construct a basis for the same.
\noindent Let $c_{\epsilon}$ be a curve in $\Or(q)$  parametrized by $\epsilon$, so that $c_{0}=q= Id \circ (q \circ Id) . \sqrt{Id'}$, where the first $Id$ is in $SO(3)$, and the second and third ones are in $\mathcal{D}$. By definition of the tangent space, $T_q{\mathcal{O}(q)}$ consists of the elements $\frac{\partial}{\partial{\epsilon}}|_{\epsilon=0}c_{\epsilon}$.\\

\noindent $\frac{\partial}{\partial{\epsilon}} |_{\epsilon=0} c_{\epsilon}$.\\
          $=\frac{\partial}{\partial{\epsilon}} |_{\epsilon=0} A_{\epsilon}(q \circ \gamma_{\epsilon}). \sqrt{\gamma_{\epsilon}'}$ \\
          $=\frac{\partial}A_{\epsilon}{\partial{\epsilon}} |_{\epsilon=0}. (q \circ \gamma_0). \sqrt{ {\gamma_0}' } + A_0 . \frac{\partial}{ \partial{\epsilon} } |_{\epsilon=0}(q \circ \gamma_{\epsilon}) .\sqrt{ {\gamma_0}' } + A_0 . (q \circ \gamma_0). \frac{\partial}{\partial{\epsilon}} |_{\epsilon=0}\sqrt{\gamma_{\epsilon}} $  (taking derivative of product of functions)\\
          $= Bq + \frac{\partial}{\partial{\epsilon}}|_{\epsilon=0} (q \circ \epsilon) + q. \frac{\partial}{\partial{\epsilon}} |_{\epsilon=0}\sqrt{\gamma_{\epsilon}} $, where $B\in Lie(SO(3)$\\
          $= Bq + (q' \circ \gamma_0) \frac{\partial}{\partial{\epsilon}}|_{\epsilon=0} {\gamma_{\epsilon}}+ q .\frac{\partial}{\partial{\epsilon}}|_{\epsilon=0}\sqrt{  \frac{\partial}{\partial{t}} \gamma_{\epsilon} (t) } $\\
          $=Bq + q'. g + q. \frac{1}{2\sqrt{ {\gamma_0}}' } \frac{\partial}{\partial{t}} \frac{\partial}{\partial{\epsilon}}|_{\epsilon=0} \gamma_{\epsilon}(t)$ (where $g \in Lie(\Diff)$)\\
          $=Bq + q' g + \frac{1}{2}q g'$\\

\begin{algorithm}
\caption{Construction of the data matrix for discretized shapes }
\label{algo:datamatrix}
\begin{algorithmic}[1]
\\ Find the K{\"a}rcher mean $\mu$ for a set of shapes $\{ q_i \}, i = 1, \ldots, N$ using Eqn. \ref{eq:karcher}. Find the corresponding tangent vectors from $\mu$  to all $q_i$s be $\{v_i \}$ following Sec. \ref{sec:mean_shape}.
\\ Let $B$ be a discrete $d$-dimensional orthogonalized Schauder basis of $\Ltwo$, where $B \equiv \{  b_i\}, i = 1, \ldots, d$
\\ Project $B$ on the tangent space $T_\mu({\cal C}^o)$ by $\tilde{B} = \{  \tilde{b}_i \}$, where $\tilde{b}_i = b_i - \langle b_i, \mu \rangle \mu$,  $i=1, \ldots, d$.
\\ Orthogonalize $\tilde{B}$ w.r.t. the inner product given in Sec. \ref{sec:srvf} as $\tilde{B} = \mbox{orth}(\tilde{B})$.
\\ Let $G_\mu$ be a discrete $k-$dimensional orthogonal basis of $T_{\mu}{\mathcal{O}(\mu)}$ from Sec. \ref{sec:fourier}.
\\ Construct the basis of $T_{\mu}({\cal S})$ as $G$, where $G \equiv \{ g_i\}$, $g_i = \tilde{b}_i - \sum_{j=1}^k \langle \tilde{b}_i, g_j \rangle g_j$, $i = 1 \ldots d$.
\\Finally construct the data matrix as $a_{ij} = \langle v_i, g_j\rangle$, for $i = 1 \ldots d, j = 1 \ldots, N$.
\end{algorithmic}
\end{algorithm}

\begin{algorithm}
\caption{Algorithm for computing Decision Boundary in QDA/LDA on shape data}
\label{expoalgorithm}
\begin{algorithmic}[1]

\\Represent each shape for each training set as a $3\times (m+1)$ matrix. 
\\Call the two classes $A$ and $B$. Let $X:= A \cup B$.
\\Compute the K{\"a}rcher mean $\mu=\mu_X$ of all the matrices (shapes) of $X$ using the Riemannian metric structure of $S$, the shape space containing all the data. This K{\"a}rcher mean is also a $3\times (m+1)$ matrix.
\\Lift each data $A_i$ in $A$ and each data $B_i$ in $B$ to the tangent space $T_{\mu_X}S$ by the inverse of the exponential map from an open subset of $\Tms$ to $S$. Call the lifting map the \emph{logarithm map}. To do this, find  the geodesics from $\mu_X$ to each $A_i$ in $A$ and each $B_i$ in $B$ in the shape space $S$, and compute the corresponding velocity vectors in $ T_{\mu_X}S \equiv \Tms $. These vectors have the same dimension for each data, i.e. each velocity vector is a $3 \times (m+1)$ matrix.
\\By slight abuse of notation, call the lift of each $A_i, B_i\in S$ to $T_{\mu_X}S$, as $A_i, B_i$ again. For the rest of the algorithm all $A_i, B_i$ will lie only in $V:=T_{\mu_X}S$.
\\(Schauder basis construction for $T_q{\Or(q)}$): Following the discussion in section \ref{sec:fourier}, compute the Schauder basis $\{g_i\}$ for $T_q{\Or(q)}$. 
\\(Orthonormal Schauder basis or Fourier basis construction for $V$): Following the discussion in section 3.2, 3.3, compute the Fourier basis $\{w_k\}$ for $\Tms$.
\\ Use Algorithm \ref{algo:datamatrix} to construct the two data matrices $X_A, X_B$ for $A, B$ respectively, having dimensions $K \times N_1, K \times N_2$ respectively.

\\(Construction of sample means) Determine the means $m_A, m_B$ of classes $A, B$ as by averaging the columns of $X_A, X_B$ respectively. Define $\tilde{m_A}:=m_A(1, 1,...1), \tilde{m_B}:=m_B(1, 1,...1)$.
\\(Construction of sample covariance matrices for discretized shape data in classes $A$ and $B$:) Form the corresponding $K \times K$ covariance matrices $\Sigma_A=\frac{1}{N_1 - 1} ({X_A - \tilde{m_A}})({X_A - \tilde{m_A}})^{T}, \Sigma_B:=\frac{1}{N_2 - 1} ({X_B - \tilde{m_B}})({X_B - \tilde{m_B}})^{T}$.

\\(Estimation of prior probability of classes) $\pi_A:=\frac{N_1}{N},\pi_B:=\frac{N_2}{N}$.

\\(Quadratic/Linear discriminant functions) Compute:  
$\delta_A(x):= -0.5 ln (det \Sigma_A) -0.5 (x-{m_A})^{T}\Sigma_A^{-1}(x-{m_A}) + ln (\pi_A):\mathbb{R}^{K}\to \mathbb{R}$ and
$\delta_B(x):= -0.5 ln (det \Sigma_B) -0.5 (x-{m_B})^{T}\Sigma_B^{-1}(x-{m_B}) + ln (\pi_B):\mathbb{R}^{K}\to \mathbb{R}$. \label{algo:line_qda_lda}
\\(Quadratic/Linear Decision Boundary) Compute the set: $ \{x\in \mathbb{R}^{K}:\delta_A(x)= \delta_B(x) \}$. \label{algo:line_lda}
\end{algorithmic}
\end{algorithm}

\section{Classification problem of shape data: Linear and Quadratic Classifiers of Shapes}
\label{sec:classifier}
Data classification problems fall into the category of a supervised machine learning problems, where we construct classifiers from the given training data, to determine the classification of a (future) test dataset, based on the previously obtained decision boundary of the classifier. One way to perform the above classification is to construct a discriminant classifier using the Bayesian {\`a} posterior probability and normality assumption on data according to the central limit theorem. The discriminant classifier is quadratic in general, and becomes linear if the two given classes of data have equal covariance matrices. Keeping this in mind, when the two data sets have almost equal covariances, we perform linear discriminant analysis (LDA) instead of quadratic discriminant analysis(QDA).

\noindent In standard settings, the discriminant analysis is performed when the data lie in some Euclidean space ${\cl}^d$ with its canonical inner product. However, it goes without saying that the theory is not straightforward for shape-valued data, since i) shape data generally do not lie in vector spaces, and ii) even after using the tangent spaces of the shape spaces to linearize the problem and after truncation of the tangent spaces to make the problem finite dimensional, the inner product of these tangent spaces mostly have non-Euclidean inner products. Here is where the construction of orthonormal Schauder/Fourier basis (Sec. \ref{sec:motivation_fourier}) comes in, which reduces the problem to one in a finite dimensional Euclidean space, following the discussion of the subsections \ref{sec:mean_discrete}, \ref{sec:mean_finite}, and \ref{sec:fourier}. This allows us to perform standard machine learning techniques on the shape data.

\subsection{Algorithmic construction of decision boundary for two classes of shape data}

Before performing discriminant analysis, we would like to represent our shapes as features in a matrix, also known as the data matrix. This is achieved as follows. Given a population of $N$ shapes, a mean shape $\mu$ and tangent vectors from the mean shape to each of the shapes of the population are computed. Then a tangent space at $\mu$ is formed and approximated using an orthonormal  Schauder basis.  Each tangent vector observation is then projected into this basis space. The coefficients of the projections are used to construct the data matrix. Algorithm \ref{algo:datamatrix} shows a procedure to construct a data matrix for a collection of shapes. 

We finally outline  the procedure  for QDA and LDA to compute the decision boundaries for the two classes in Algorithm \ref{expoalgorithm}. Our algorithm is a generic one that can be used for both QDA and LDA, since in step \ref{algo:line_qda_lda}, the quadratic terms cancel out for the case of equal covariance matrices, producing a linear decision boundary in step \ref{algo:line_lda}. 

\section{Data, Experiments, and Results}
\label{sec:results}
In this section we first present results on synthetic data and subsequently show classification results on data from three examples, i) cortical sulcal shapes, ii) a population of corpus callosum shapes of healthy controls and patients with schizophrenia, and iii) facial midline profiles for healthy individuals and patients with severe alcohol exposure from a fetal alcohol syndrome cohort.   

\subsection{Synthetic Data}
\begin{figure}[htb]
\centering
\includegraphics[width=\columnwidth]{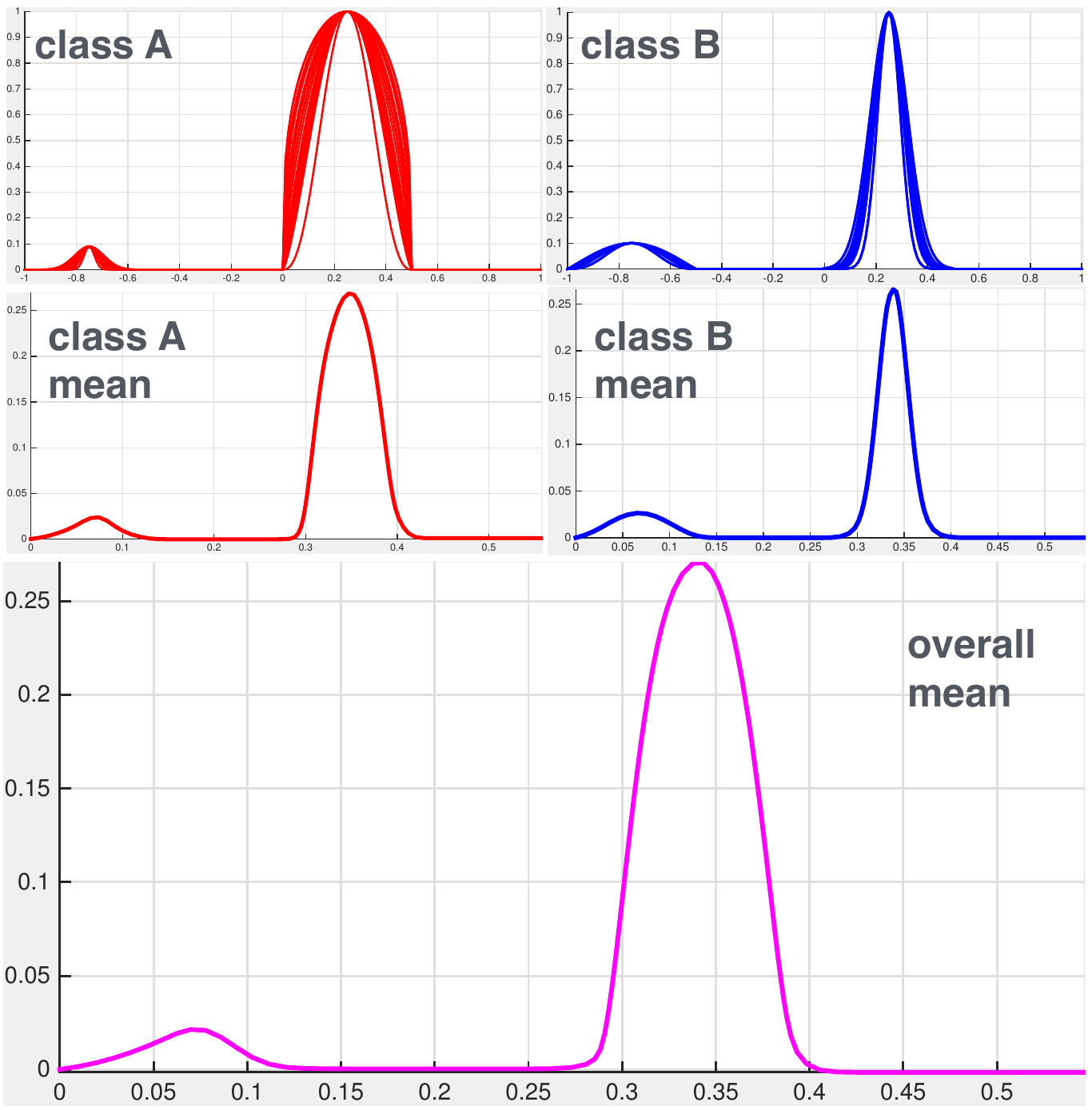}
\caption{Top row: 100 samples each from shapes of class A and B. Middle row: Mean shapes for the two classes. Bottom row: mean shape of the population.}
\label{fig:synthetic}
\end{figure}

Figure \ref{fig:synthetic}  shows examples of two classes of synthetic shape data consisting of two bumps (gaussians followed and preceded by powers of half-sinusoids for the two classes respectively). The variation is obtained by changing the pulse width of the gaussians and powers of sinusoids. 100 shapes each of class A and B were generated and their mean shape was computed on the shape space. 10-fold repeated cross validation was performed to test a quadratic classifier. The average cross validation error was obtained as $0.031$ (accuracy of 96.9\%).

\subsection{Classification of Cortical Sulci}

\begin{figure*}[htb]
\centering
\includegraphics[width=\textwidth]{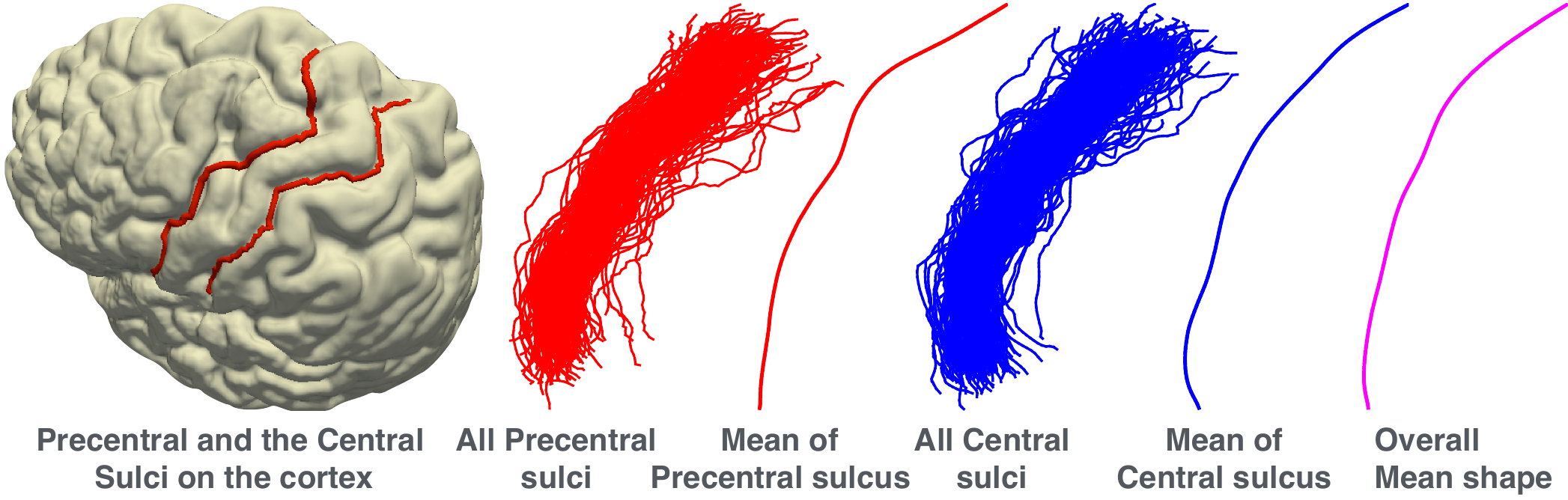}
\caption{Left to right: examples of the precentral and the central sulci on the cortex, population of all 175 precentral sulci, mean shape of the precentral sulci, population of all 175 central sulci, mean shape of the central sulci, and the mean shape of the overall population. }
\label{fig:sulci}
\end{figure*}

Here we show results of classification of cortical sulci in a population of  175 human subjects from ages 18 to 50 years. Subjects's brains were scanned on a 1.5T MRI Siemens scanner with a 1mm$^3$ isotropic voxel resolution.  The MRI images underwent skull stripping and tissue editing, and the gray matter boundary was  extracted \cite{holmes1996cortical} to obtain triangular mesh representations of the cortex.  A set of 26 landmark sulci were traced on the cortical surface by an expert neuroanatomist.  Since this paper is focused on discriminant analysis, for the purpose of demonstration, we chose to show results of binary classification for the central and the precentral sulci as shown in Figure \ref{fig:sulci}. Sulcal landmarks were represented as open curves   and approximated by the tangent space Fourier representation at the mean shape for both the classes. 

Figure \ref{fig:sulci} shows an example of the precentral and the central sulci traced on an individual subject,  the population of the precentral and the central sulci and their respective means, as well as the overall mean of the full dataset. We used a  repeated 10-fold cross validation to evaluate the a linear discriminant classifier to obtain an average cross validation error of 12.75\% (accuracy of 87.25 \%) with a model training error of 5.72\%. 

\subsection{Classification of Schizophrenia Diagnostic Status using Corpus Callosum Shapes}
\begin{figure}[h]
\centering
\includegraphics[width=0.5\textwidth]{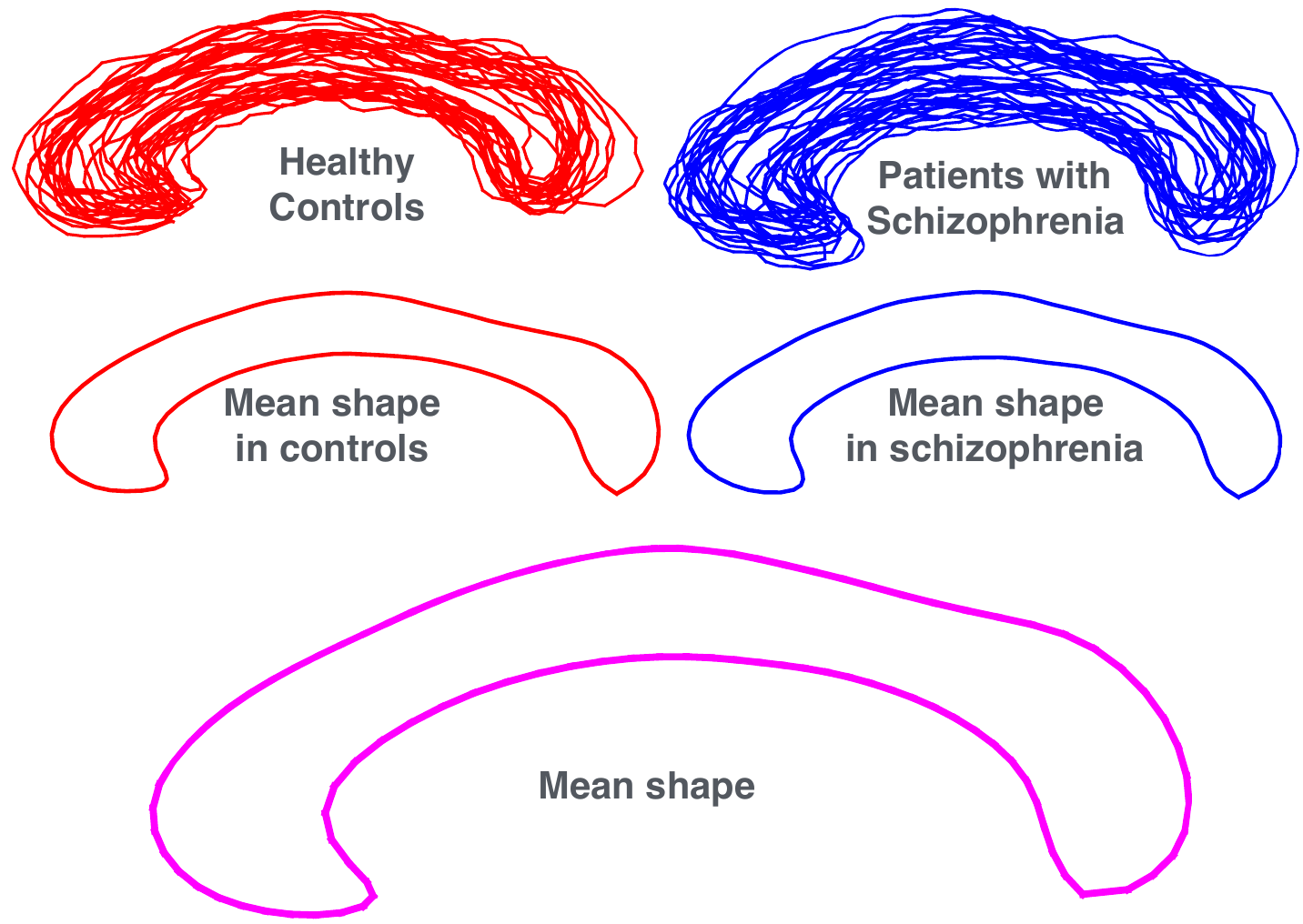}
\caption{Top row: 20 corpus callosum shapes each for both healthy controls and patients with schizophrenia. Middle row: Mean shapes for the healthy controls and schizophrenia. Bottom row: mean shape of the population.}
\label{fig:CC}
\end{figure}

Emerging evidence \cite{roy2007loss} suggests the important role of white matter abnormalities  in schizophrenia. Consequently  one expects disease effects on callosal morphology in patients. Here we show results of classification of schizophrenia diagnosis in a population of  40 subjects (20 healthy controls and 20 patients with schizophrenia) from ages 18 to 46 years. MRI images for all subjects were acquired on a 1.5T Siemens scanner with 1mm$^3$ isotropic voxel resolution. An expert neuroanatomist traced the callosal boundary on the mid sagittal slice for each subject. The boundaries were represented in the shape space and the tangent space Fourier approximation was obtained for each callosal shape. Figure \ref{fig:CC} shows the callosal shapes for controls and patients. After a repeated 10-fold cross validation to evaluate the linear classifier, we obtained an average cross validation error of 14.67\% (accuracy of 85.33 \%) with a model training error of 5\%. 

\subsection{Classification of Severe Alcohol Exposure using Facial Midline Curves}
\begin{figure}[h]
\centering
\includegraphics[width=0.5\textwidth]{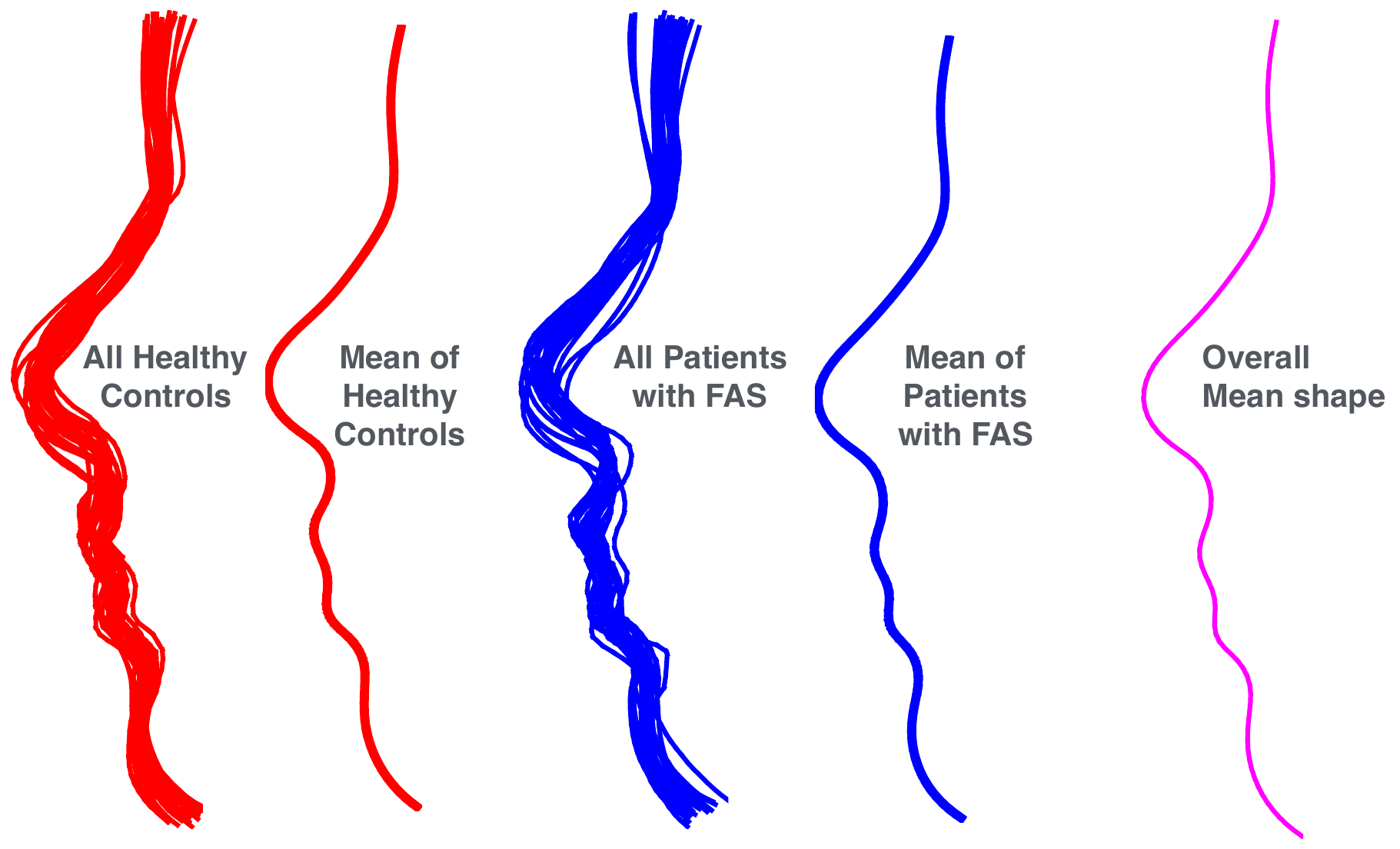}
\caption{Left to right: midline facial profile curves for 24 healthy controls, mean shape of  healthy controls, 21 facial profile curves for patients with FAS, mean shape of patients, and the mean shape of the overall population.}
\label{fig:FAS}
\end{figure}

Fetal alcohol spectrum disorders arise from moderate to heavy prenatal alcohol exposure to the fetus during development of the central nervous system (CNS).  Fetal alcohol syndrome (FAS), which is a more severe case of FASD is usually characterized  identifiable facial dysmorphology \cite{Suttiee779}, growth deficits, and behavioral problems. Children and adolescents diagnosed with FAS exhibit a  variety of minor to severe facial abnormalities including flat nasal bridge, smooth philtrum (space between nasal base and upper lip), short palpebral fissure (space between eyelids), thin upper lip, and a flat mid-face \cite{jones1973recognition}. Here we attempt to classify patients with FAS from healthy controls using the midline curve from 3D facial surfaces. Our data consisted of 45 subjects (21 with FAS and 24 healthy controls) from ages 5 to 16 years. 3D facial range scans were acquired using a Minolta Vivid 910 laser scanner (1/3 frame CCD 340K pixels, triangulation light block method, scan range- 0.6 - 2.5 m, 0.3 sec. peak acquisition time).  The facial midline curve was automatically extracted as an intersection of a coronal center plane passing through the nose and the facial surface. Shapes of all curves were represented by their SRVFs and a tangent space was constructed at the mean shape as shown in section \ref{sec:mean_shape}. Figure \ref{fig:FAS} shows the facial midline shapes for healthy controls and patients with FAS, and the mean shape of the population. A linear discriminant  classifier was trained on this data and a repeated 10-fold cross validation was performed. This data yielded an average cross validation error of 0.2378 (accuracy of 76.22\%) with a training error of 8\%. 

Finally, all the results are summarized in a tabular form in Table \ref{tab:performance}.

\begin{table}[t]
\label{tab:performance}
\centering
    \caption{Classifier Performance for different types of data.}
    \begin{tabular}{l l l l l  }
    \toprule 
       Example & Classifier & N (size) & \multicolumn{1}{p{1cm}}{\centering Training \\ Accuracy } & \multicolumn{1}{p{1cm}}{\centering Testing \\ Accuracy } \\ \midrule
       \rowcolor[gray]{.9} 
      toy & qda & 200  & 100\% & 96.9\%  \\ 
      sulci & lda & 175  & 94.28\% & 87.25\%  \\ 
        \rowcolor[gray]{.9} 
        corpus callosum & lda & 40  & 95\% & 85.33\% \\ 
 	facial curves & lda & 45  & 92\% & 76.22\% \\ 
         \bottomrule
    \end{tabular}
\end{table}

\section{Discussion}
We presented a geometric approach for defining classifiers on the shape space of curves. In this work, we focused on quadratic and linear discriminant analysis, although the approach is general and  allows for defining more advanced classification techniques on the shape space.  While we obtained high accuracy on synthetic data as well as corpus callosum shapes, we obtained a moderate accuracy on  FAS facial dysmorphology data. This could be explained by the fact that the FAS data consisted of just the midline facial curves instead of the whole face. We expect the accuracy to improve after combining the shape classifier with anthropomorphic measures such as the palpebral fissure spans or the philtrum lengths. 

Future work will focus on implementing  more advanced machine learning approaches for shapes using the geometric framework proposed in this paper. 

\section{Acknowledgment}
This research was supported by National Institute on Alcohol Abuse and Alcoholism  award K25 AA024192.

{\small
\bibliographystyle{ieee}
\bibliography{paper}
}

\end{document}